\documentstyle[aps,preprint,prl]{revtex}
\begin{document}
\flushbottom

\widetext
\draft
\title{Some exact analytical results and a semi-empirical formula for single
electron ionization induced by ultrarelativistic heavy ions}
\author{A. J. Baltz}
\address{
Physics Department,
Brookhaven National Laboratory,
Upton, New York 11973}
\date{\today}
\maketitle

\def\thepage{\arabic{page}}
\makeatletter
\global\@specialpagefalse
\ifnum\c@page=1
\def\@oddhead{Draft\hfill To be submitted to Phys. Rev. A}
\else
\def\@oddhead{\hfill}
\fi
\let\@evenhead\@oddhead
\def\@oddfoot{\reset@font\rm\hfill \thepage \hfill}
\let\@evenfoot\@oddfoot
\makeatother

\begin{abstract}
The delta function gauge of the electromagnetic potential allows semiclassical
formulas to be obtained for the probability of exciting a single electron out
of the ground state in an ultrarelativistic heavy ion reaction.  Exact
formulas have been obtained in the limits of zero impact parameter and large,
perturbative, impact parameter.  The perturbative impact parameter result can
be exploited to obtain a semi-empirical cross section formula of the form
$\sigma = A \ln \gamma + B$ for single electron ionization.  $A$ and $B$ can be
evaluated for any combination of target and projectile, and the resulting
simple formula is good at all ultrarelativistic energies.  The analytical
form of $A$ and $B$ elucidates a result previously found in numerical
calculations: scaled ionization cross sections decrease with increasing charge
of the nucleus being ionized.  The cross section values obtained from the
present formula are in good agreement with recent CERN SPS data
from a Pb beam on various nuclear targets.
\end{abstract}

\makeatletter
\global\@specialpagefalse
\def\@oddhead{\hfill}
\let\@evenhead\@oddhead
\makeatother
\nopagebreak
\section{Introduction}
In a recent work\cite{ajb2} ionization cross sections were calculalated for a
number of representative cases of collisions involving ultrarelativistic
Pb, Zr, Ca, Ne and H ions.  The method of calculation (on a computer) involved
an exact semiclassical solution of the Dirac equation in the
ultrarelativistic limit\cite{ajb1}.  A single electron was
taken to be bound to one nucleus with the other nucleus completely
stripped.  The probability that the electron would be ionized in the
collision was calculated as a function of impact
parameter, and cross sections were then constructed by the usual integration 
of the probabilities over the impact parameter.  The results of the probability
calculations were used to construct cross sections for various ion-ion
collision combinations in the form
\begin{equation}
\sigma =  A \ln \gamma + B
\end{equation}
where $A$ and $B$ are constants for a given ion-ion pair and
$\gamma\ (=1/\sqrt{1-v^2})$ is the
relativistic factor one of the ions seen from the rest frame of the other.

In Section II of this paper analytic results are derived for the
probability that a single ground state electron will be excited in an
ultrarelativistic heavy ion reaction.  Exact semiclassical formulas are
presented for the limits of zero impact
parameter and perturbational impact parameters.  In Section III the
perturbational impact parameter analytical form is used as a basis
to construct semi-empirical formulas for $A$ and $B$. These formulas
reproduce the previous numerical results for single particle ionization,
and they illuminate the systematic
behavior of $A$ and $B$ with changing target and projectile ion species.
Ionization cross sections calculated with Eq.(1) are then compared with data.

\section{Impact parameter dependent probabilities}
If one works in the appropriate gauge \cite{brw}, then
the Coulomb potential produced by an ultrarelativistic particle (such as a
heavy ion) in uniform motion can be expressed in the following form\cite{ajb}
\begin{equation}
V(\mbox{\boldmath $ \rho$},z,t)
=-\alpha Z_1 (1-\alpha_z) \delta(z - t)
\ln{({\bf b}-\mbox{\boldmath $ \rho$})^2 \over b^2 }.
\end{equation}
${\bf b}$ is the impact parameter, perpendicular to the $z$--axis along which
the ion travels, $\mbox{\boldmath $\rho$}$, $z$, and $t$ are the coordinates of
the potential relative to a fixed target (or ion),
$\alpha_z$ is the Dirac matrix, $\alpha$ 
the fine structure constant, and
$Z_1$ and $v$ the charge and velocity of the moving ion.
This is the physically relevant ultrarelativistic potential since it was 
obtained by ignoring terms in
$({\bf b} - \mbox{\boldmath $\rho$}) / \gamma^2$\cite{ajb}\cite{brw}.
Its multipole expansion is
\begin{eqnarray}
V(\mbox{\boldmath $ \rho$},z,t)&=&\alpha Z_1 (1-\alpha_z) 
\delta (z - t) \nonumber \\
&& \biggl\{ -\ln {\rho^2 \over b^2}\ \ \ \ \ \ \ \ \ \ \ \ \  \rho>b\nonumber\\
&&+\sum_{m>0}{2 \cos m \phi \over m}\nonumber\\
&&\times\biggl[\biggl({\rho \over b}\biggr)^m
 \ \ \ \ \ \rho <b \nonumber\\
&&+\biggl({b \over \rho}\biggr)^m\biggr]\biggr\}.
 \ \ \ \ \ \rho >b
\end{eqnarray}
For $ b >> \rho $
\begin{equation}
V(\mbox{\boldmath $ \rho$},z,t)
=\delta (z - t) \alpha Z_1(1-\alpha_z) 
2 {\rho \over b} \cos \phi .
\end{equation}
  As will
be shown in Section III, when
${\bf b}$ becomes large enough that expression Eq.(4) is inaccurate for use
in calculating a probability, we
match onto a Weizsacker-Williams expression which is valid for large
$b$.  Note that the $b^2$ in the denominator of the logarithm in Eq.(2) is
removable by a gauge transformation, and we retain the option of keeping or
removing it as convenient.

It was shown in Ref. \cite{ajb1} that the $\delta$ function allows the
Dirac equation to be solved exactly at the point of interaction, $z = t$.
Exact amplitudes then take the form
\begin{eqnarray}
a_f^j(t=\infty)&=\delta_{fj} & +
\int_{-\infty}^{\infty} d t e^{i (E_f - E_j) t} \langle \phi_f \vert
\delta(z - t) (1-\alpha_z) \nonumber \\ 
&&\times ( e^{-i \alpha Z_1 \ln{({\bf b}- \mbox{\boldmath $ \rho$})^2 }} - 1 ) 
\vert \phi_j \rangle
\end{eqnarray}
where $j$ is the initial state and $f$ the final state.  This amplitude is in
the same form as the perturbation theory amplitude, but with
an effective potential to represent all the higher order effects exactly,
\begin{equation}
V(\mbox{\boldmath $ \rho$},z,t)
=-i \delta(z - t) (1-\alpha_z)
( e^{-i \alpha Z_1 \ln{({\bf b}- \mbox{\boldmath $ \rho$})^2 }} - 1 ) ,
\end{equation}
in place of the potential of Eq.(2).  

Since an exact solution must be unitary,
the ionization probability (the sum of probabilities of excitation from the
single bound electron to particular continuum states) is equal
to the deficit of the final bound state electron population
\begin{equation}
\sum_{ion} P(b) = 1 - \sum_{bound} P(b)
\end{equation}
The sum of bound state probabilities includes the probability that the electron
remains in the ground state plus the sum of probabilities that it ends up in an
excited bound state.
From
Eq.(5) one may obtain in simple form the exact survival probability of an
initial state
\begin{equation}
P_j (b) =  \vert \langle \phi_j \vert (1-\alpha_z)
e^{-i \alpha Z_1 \ln{({\bf b}- \mbox{\boldmath $ \rho$})^2 }} 
\vert \phi_j \rangle \vert^2.
\end{equation}
By symmetry the $\alpha_z$ term falls out and we are left with
\begin{equation}
P_j (b) =  \vert \langle \phi_j \vert
e^{-i \alpha Z_1 \ln{({\bf b}- \mbox{\boldmath $ \rho$})^2 }} 
\vert \phi_j \rangle \vert^2.
\end{equation}
The ground state wave function $ \phi_j $ is the usual K shell Dirac
spinor\cite{rose}
\begin{equation}
\phi_j = \left( \begin{array}{c} g(r) 
\chi_{\kappa}^{\mu} \\ i f(r) \chi_{-\kappa}^{\mu} \end{array} \right)
\end{equation}
with upper and lower components wave functions $g$ and $f$ 
\begin{eqnarray}
g(r) & = & N \sqrt{1 + \gamma_2}\ r^{\gamma_2-1}\ e^{-\alpha Z_2 r} \nonumber 
\\
f(r) & = & -N \sqrt{1 - \gamma_2}\ r^{\gamma_2-1}\ e^{-\alpha Z_2 r}
\end{eqnarray}
where $Z_2$, is the charge of
the nucleus that the electron is bound to,
$\gamma_2 = \sqrt{1 - \alpha^2 Z_2^2}$, and
\begin{equation}
N^2 = {(2 \alpha Z_2)^{2\gamma_2 + 1}\over 2 \Gamma(\gamma_2 + 1)}.  
\end{equation}

Let us first consider $b = 0$.  We have
\begin{equation}
P_j (b=0) =  \vert \langle \phi_j \vert
e^{-2 i \alpha Z_1 \ln{\rho}} 
\vert \phi_j \rangle \vert^2 =  \vert \langle \phi_j \vert
e^{-2 i \alpha Z_1 (\ln{r} + \ln(\sin{\theta}))}
\vert \phi_j \rangle \vert^2.
\end{equation}
Putting in the explicit form of the upper and lower components for the K shell
lowest bound state Dirac wave function and carrying out the integration we have
\begin{equation}
P_j (b=0) = {\pi \over 4} \Bigg\vert {\Gamma (2 \gamma_2 + 1 - 2 i \alpha Z_1)
\Gamma (1 - i \alpha Z_1) \over \Gamma (2 \gamma_2 + 1) \Gamma({3 \over 2} -
i \alpha Z_1)} \Bigg\vert^2,
\end{equation}
or
\begin{equation}
P_j (b=0) = {\pi \alpha Z_1 \, {\rm ctnh} (\pi \alpha Z_1) \over 
(1 + 4  \alpha^2 Z_1^2)} \Bigg\vert {\Gamma (2 \gamma_2 + 1 - 2 i \alpha Z_1)
 \over \Gamma (2 \gamma_2 + 1) } \Bigg\vert^2.
\end{equation}
It is interesting to compare this result with a previous calculation of the
probability of ionization in ``close collisions''  by Bertulani and
Baur\cite{bb}.  For a one electron atom they find
\begin{equation}
P_{ion} (b < \lambda_c / \alpha Z_2) = 1.8 \ \alpha^2 Z_1^2,
\end{equation}
where $\lambda_c = \hbar / m_e c$ is the electron Compton wavelength.
If we take the low $Z_1$ limit of our expression Eq.(15) and then subtract it
from one we obtain 
\begin{equation}
P_{ion} (b = 0) = ({\pi^2 \over 3} -1 )\alpha^2 Z_1^2 = 2.29 \  \alpha^2 Z_1^2
\end{equation}
However our expression Eq.(15) only gives the flux lost from the initial state;
some of that flux goes into excited bound states and is not ionized.  From our
previous numerical calculations we find that the actual ionization
probabilities obtained either by summing up final continuum states or else by
subtracting all the final bound states from unity were 76\% -- 80\%
respectively of the flux lost from the initial state.  Thus if
we multiply the constant in Eq.(17) by such a
 percentage
we are in remarkable agreement with Bertulani and
Baur for the perturbative limit.

Now let us consider the case of $b >> \rho$.  From Eq.(4) and Eq.(9) we have
\begin{equation}
P_j (b) =  \vert \langle \phi_j \vert
e^{-2 i \alpha Z_1 \cos(\phi) (\rho / b)}
\vert \phi_j \rangle \vert^2.
\end{equation}
Expanding the exponential up to $\rho^2 / b^2$ we have
\begin{equation}
P_j (b) =  \vert \langle \phi_j \vert
1 {-2 i \alpha Z_1 \cos(\phi) {\rho \over b}}
{-2 \alpha^2 Z_1^2 \cos^2(\phi) {\rho^2 \over b^2}}
\vert \phi_j \rangle \vert^2
\end{equation}
The term in $\cos(\phi)$ vanishes by symmetry, and integrating, we obtain
\begin{equation}
P_j(b) = 1 - 2 {Z_1^2 \over Z_2^2} {(1 + 3 \gamma_2 + 2 \gamma_2^2) \over 3}
{\lambda_c^2 \over b^2}
\end{equation}
by ignoring the term in $1/b^4$.

Both limits, Eq.(15) for $b = 0$ and Eq.(20) for $b >> \rho$, are
relativistically correct and thus correct for all $Z_1$ and $Z_2$ since
exact Dirac wave functions were used.

\section{A semi-empirical formula for single electron ionization}

It is well known that the cross section for ionization of any pair of
projectile and target species can
be expressed as a sum of a constant term and a term going as the log of the
relativistic $\gamma$ of the beam as seen in the target rest
frame\cite{bb}\cite{ab}\cite{ajb2}
\begin{equation}
\sigma_{ion} =  A \ln \gamma + B.
\end{equation}
The cross section of this form is constructed from an impact parameter
integral
\begin{equation}
\sigma_{ion}  = 2 \pi \int P(b)_{ion}\ b\ d b 
\end{equation}
where $P(b)$ is the probability of ionization at a given impact parameter.
If all the flux lost from the initial state went into the continuum then
Eq.(20) would provide the ionization probability at moderately large $b$
\begin{equation}
P_{ion}(b) = 2 {Z_1^2 \over Z_2^2} {(1 + 3 \gamma_2 + 2 \gamma_2^2) \over 3}
{\lambda_c^2 \over b^2}.
\end{equation}

We will take this form as a physical basis to build a semi-empirical formula
for ionization.  In any case we need to integrate the
probability up to a natural energy cutoff.  In order to do this we match
the delta function solution Eq.(23) at some moderately large $b$ onto 
the known Weizsacker-Williams probability for larger
$b$ by noting that if $b \omega << \gamma$ then
\begin{equation}
K_1^2({\omega b \over \gamma}) = {\gamma^2 \over \omega^2 b^2},
\end{equation}
and we can rewrite Eq.(24) in the Weizsacker-Williams form for large $b$
\begin{equation}
P_{ion}(b) = 2 {Z_1^2 \over Z_2^2} {(1 + 3 \gamma_2 + 2 \gamma_2^2) \over 3}
{\lambda_c^2 \omega^2 \over \gamma^2} K_1^2({\omega b \over \gamma}).
\end{equation}
To perform the large $b$ cutoff recall that to high degree of accuracy
\begin{equation}
{\omega^2 \over \gamma^2} \int_{b_0}^\infty K_1^2({\omega b \over \gamma})
b\ db = \ln({0.681 \gamma \over \omega b_0}) = 
\ln \gamma  + \ln({0.681 \over \omega b_0}).
\end{equation}
We immediately obtain
the following expression for $A$
\begin{equation}
A = {4 \pi \lambda_c^2 \over 3} {Z_1^2 \over Z_2^2} (1 + 3 \gamma_2
+ 2 \gamma_2^2),
\end{equation}
where $\lambda_c^2$, the square of the electron Compton wave length,
is 1491 barns.
However, as it turns out, uniformly for all species of heavy ion reactions,
at perturbational impact parameters a little over $70\%$ of the flux lost 
from the initial state goes into excited
bound states and does not contribute to ionization.  But since the ratio of
flux going into continuum states to the total flux lost is so uniform we can
use a fit to previously published numerical results\cite{ajb2} to obtain a
semi-analytical form for $A$:
\begin{equation}
A = (0.2869) {4 \pi \lambda_c^2 \over 3}{Z_1^2 \over Z_2^2} 
(1 + 3 \gamma_2 +2 \gamma_2^2),
\end{equation}
or in barns
\begin{equation}
A = 1792 {Z_1^2 \over Z_2^2} (1 + 3 \gamma_2 +2 \gamma_2^2).
\end{equation}

Now one can use the second term in Eq.(26) to obtain a provisional expression
for $B$
\begin{equation}
B = A \ln({0.681 \over \omega b_0}).
\end{equation}
Obviously we need to evaluate $\omega$ and to discuss $b_0$.  $\omega$ can be
taken as the minimum ionization energy, $1 - \gamma_2$,
times a constant a little larger than one.
One next observes that if $P_{ion}(b)$ varies as $1/b^2$
the impact parameter integral has to be cut off on the low side
at some value $b_0$ to
avoid divergence.  In fact the $1/b^2$ dependence continues down to the
surface of the atom where other terms evident in Eq.(3) begin to contribute.
The atomic size is just the electron Compton wave length divided by
$\alpha Z_2$.  In this region $P_{ion}(b)$ first rises faster than $1/b^2$
and then levels off to approach a constant change with $b$ at
$b=0$\cite{ajb2}.  One could try to add a low impact paramenter contribution
to $A$ based on Eq.(15) to our provisional form Eq.(30), but that turns out
to unduly
complicate things without improving the phenomenology.  Our approach will
be to set $b_0$ to an empirical constant divided by $\alpha Z_2$

Eq.(30) now takes the form
\begin{equation}
B = A \ln ({C \alpha Z_2
\over 1 - \gamma_2}).
\end{equation}
Putting in two analytical fine tuning factors and fitting the remaining
constant to
the numerical results of Ref.\cite{ajb2} we obtain a  
semi-analytical form for $B$:
\begin{equation}
B = A \gamma_1^{1/10} (1 - \alpha^2 Z_1 Z_2)^{1/4} \ln ({2.37 \alpha Z_2
\over 1 - \gamma_2}).
\end{equation}

Table I expands a corresponding table from Ref.\cite{ajb2} by adding cross
sections of symmetric ion-ion pairs calculated with the formulas for $A$ and
$B$.  There is good agreement between the formula values for the cross sections
(first rows) and the numerical cross sections calculated by subtracting the
bound state
probabilities from unity (second rows) or calculated by summing continuum
electron final states (third rows).  For both $A$ and $B$ the agreement is
also good with the Anholt and Becker
calculations\cite{ab} in the literature for the lighter ion species.
However with increasing mass of the ions the perturbative energy dependent
term $A$ decreases in the formula calculations and in our previous numerical
calculations, whereas it increases in the
Anholt and Becker calculations.  The greatest discrepancy is for Pb + Pb,
with Anholt and Becker being about 60\% higher.  The reason that the $A$
should decrease with increasing mass (actually $Z$) of the ions
is explained by the 
\begin{equation}
(1 + 3 \gamma_2 +2 \gamma_2^2) = 3 - 2 \alpha^2 Z_2^2 + 3 
\sqrt {1 - \alpha^2 Z_2^2}
\end{equation} 
factor in the formula for $A$ (and thereby $B$ also).
As we noted before, perhaps the discrepancy between our $A$ decreasing with
$Z$ and the Anholt and Becker $A$ increasing with $Z$
is due to the fact that Anholt and Becker use approximate relativistic bound
state wave functions and the present calculations utilize exact
Dirac wave functions for the bound states.  
For the term
$B$ (which has the non-perturbative component) the agreement is relatively
good between all the calculations.

In the perturbative limit (small $Z_1, Z_2$) the cross section formula goes
over to
\begin{equation}
\sigma = (0.2869) 8 \pi \lambda_c^2 {Z_1^2 \over Z_2^2} 
\ln{2.37 \gamma  \over \alpha Z_2} = 7.21 \lambda_c^2 {Z_1^2 \over Z_2^2} 
\ln{2.37 \gamma  \over \alpha Z_2}.
\end{equation}
By way of comparison, Bertulani and Baur\cite{bb} using the equivalent photon
method and taking the contribution of $b \ge \lambda_c / \alpha Z_2$ found
\begin{equation}
\sigma =  4.9 \lambda_c^2 {Z_1^2 \over Z_2^2} 
\ln{2 \gamma \over \alpha Z_2}
\end{equation}
for this case of ionization of a single electron. 

Table II shows results of the calculation of $B$ (multiplied by $Z_2^2/Z_1^2$)
for a number of representative
non-symmetric ion-ion pairs.  (Since $A$ is perturbative, scaling as
$Z_1^2$, its value can be taken from Table I for the various pairs here.)
Once again there is good agreement between the formula values for the cross
sections (first rows) and the numerical cross sections calculated by
subtracting the bound state
probabilities from unity (second rows) or calculated by summing continuum
electron final states (third rows).  The only notable disagreement is with
Anholt and Becker for Pb targets.

The availabilty of the present semi-empirical formula facilitates a comparison
with available CERN SPS data.  Calculations with the formula
are in considerably better agreement with the data of 
Krause et al.\cite{kr} for a Pb beam on various targets than are the Anholt and
Becker numbers with or without screening.  Note that in this case the role
of target and beam are reversed.  It is the single electron Pb ion in the beam
that is ionized by the various nuclei in the fixed targets.  The formula
numbers do not include screening, which should in principle be included for a
fixed target case.  However, one might infer from the Anholt and Becker
calculations that the effect of screening is smaller than the error induced
by using an approximate rather than proper relativistic wave function for the
electron bound in Pb.  

Note that the formula has not been fit to experimental data.  It is compared
with experimental data.  The ``empirical'' aspect of this
formula refers to adjusting the formula to previous numerical calculations of
Ref.\cite{ajb2}

At RHIC the relativistic $\gamma$ of one ion seen
in the rest frame of the other is 23,000, and of course there is
no screening, so the present formula should be completely applicable.
The present formula predicts a single electron ionization cross
section of 101 kilobarns for Au + Au at RHIC.  The corresponding cross section
from Anholt and Becker is 150 kilobarns.  

\section{Acknowledgments}
After this work was completed, a paper by Voitkiv, M\"uller and
Gr\"un\cite{vmg}, which
includes screening in ionization of relativistic projectiles, was brought to my
attention.  I would like to thank Carlos Bertulani for pointing out this paper
to me and for reading the present manuscript.  This
manuscript has been authored under Contract No. DE-AC02-98CH10886 with
the U. S. Department of Energy. 
\vfill\eject

\begin{table}
\caption[Table I]{Calculated Ionization Cross Sections Expressed in the
Form $A \ln \gamma + B$ (in barns)}
\begin{tabular}{|cc|ccccc|}
&& Pb + Pb & Zr + Zr & Ca + Ca & Ne + Ne & H + H\\
\tableline
$\ \ A$&Formula& 8400 & 10,212 & 10,618 & 10,718 & 10,752 \\
&$1 - \sum_{bnd} e^-$ & 8680 & 10,240 & 10,620 & 10,730 & 10,770 \\
&$\sum_{cont} e^-$& 8450 & 9970 & 10,340 & 10,440 & 10,480 \\
&Anholt \& Becker\cite{ab}& 13,800 & 11,600 & 10,800 & 10,600 & 10,540 \\
\tableline
$\ \ B$&Formula& 14,133 & 27,375 & 36,623 & 44,638 & 69,629 \\
&$1 - \sum_{bnd} e^-$& 14,190 & 28,450 & 38,010 & 46,080 & 71,090 \\
&$\sum_{cont} e^-$& 12,920 & 27,110 & 36,530 & 44,430 & 68,780 \\
&Anholt \& Becker& 13,000 & 27,800 & 37,400 & 45,400 & 70,000\\
\end{tabular}
\label{tabi}
\end{table}
\begin{table}
\caption[Table II]{Calculated values of the scaled quantity 
$(Z_2^2 / Z_1^2) B$ for non-symmetric combinations
of colliding particles.  The second nucleus ($Z_2$) is taken to be the one with
the single electron to be ionized.  
Since Anholt and Becker cross sections without screening are completely
perturbative, their values of of B also can be taken from Table IV, and are
repeated here for convenient comparison.}
\begin{tabular}{|c|cccccc|}
& H + Ne & H + Ca & Ca + H & H + Zr & H + Pb & Pb + H \\
\tableline
Formula & 44,716& 36,890 & 69,462 & 28,226 & 16,487 & 66,539\\
$1 - \sum_{bnd} e^-$ & 46,150 & 38,270 & 70,820 & 29,440 & 17,090 & 67,550\\
$\sum_{cont} e^-$& 44,490 & 36,790 & 68,520 & 28,070 & 15,680 & 65,330\\
Anholt \& Becker\cite{ab}& 45,400 & 37,400 & 70,000 & 27,800 & 13,000 & 70,000\\
\tableline
\tableline
& Pb +Ne & Ne + Pb & Pb + Ca & Ca + Pb & Pb + Zr & Zr + Pb \\
\tableline
Formula & 42,308 & 16,313 & 34,503 & 16,097 & 25,751 & 15,592\\
$1 - \sum_{bnd} e^-$& 42,560 & 17,030 & 34,720 & 16,870 & 26,010 & 16,250\\
$\sum_{cont} e^-$& 41,000 & 15,690 & 33,330 & 15,530 & 24,730 & 14,930\\
Anholt \& Becker& 45,400 & 13,000 & 37,400 & 13,000 & 27,800 & 13,000\\
\end{tabular}
\label{tabii}
\end{table}
\begin{table}
\caption[Table III]{Cross sections for the ionization of a 160 GeV/A
one electron Pb projectile ($Z_2$) by various fixed nuclear targets ($Z_1$).
Unlike in Table II,
here the appropriate $(Z_1^2 / Z_2^2)$ factor has been included.  Cross
sections are given in kilobarns to match the format of the CERN SPS data.}
\begin{tabular}{cccccc}
Target & $Z_1$ & Formula & SPS Data & Anholt \& Becker & Anholt \& Becker\\
&&&& (with screening) & (no screening)\\
\tableline
Be & 4 & 0.14 & 0.14 & 0.24 & 0.20\\
C & 6 & 0.32 & 0.31 & 0.49 &0.45\\
Al & 13 & 1.5 & 1.3 & 2.0 & 2.1\\
Cu & 29 & 7.4 & 6.9 & 9.0 & 10.5\\
Sn & 50 & 22 & 15 & 25 & 31\\
Au & 79 & 53 & 42 & 60 & 78\\
\end{tabular}
\label{tabiii}
\end{table}
\end{document}